\documentclass[english]{article}

\usepackage[linguistics]{forest}
 \usepackage{xcolor}
 \usepackage{booktabs}
 \usepackage{colortbl}
 \usepackage{newfloat}
\usepackage[caption=false]{subfig}
 \usepackage[hyphens]{url}
\usepackage{tabularx}
\usepackage{lipsum,adjustbox}
\usepackage{algorithm,algpseudocode}
\usepackage{amsmath}
\usepackage{amssymb}
\usepackage{balance}
\usepackage{todonotes}
\usepackage{multirow}
\usepackage{footnote}
\makesavenoteenv{table}
\usepackage[multiple]{footmisc}
\begin{document}

\title{HEDGE: Efficient Traffic Classification of Encrypted and Compressed Packets}

\author{Fran Casino\\ Department of Informatics\\University of Piraeus \and Kim-Kwang Raymond Choo\\Department of Information Systems and Cyber Security\\ University of Texas at San Antonio \and Constantinos Patsakis\\ Department of Informatics\\University of Piraeus}

\maketitle

\begin{abstract}
As the size and source of network traffic increase, so does the challenge of monitoring and analysing network traffic. Therefore, sampling algorithms are often used to alleviate these scalability issues. However, the use of high entropy data streams,  through the use of either encryption or compression, further compounds the challenge as current state of the art algorithms cannot accurately and efficiently differentiate between encrypted and compressed packets. In this work, we propose a novel traffic classification method named HEDGE (High Entropy DistinGuishEr) to distinguish between compressed and encrypted traffic. HEDGE is based on the evaluation of the randomness of the data streams and can be applied to individual packets without the need to have access to the entire stream. Findings from the evaluation show that our approach outperforms current state of the art. We also make available our statistically sound dataset, based on known benchmarks, to the wider research community.
\end{abstract}

\section{Introduction}
\label{sec:intro}

Network traffic is increasing at an unprecedented race. For example, studies from IBM\footnote{\url{https://www.ibm.com/blogs/insights-on-business/consumer-products/2-5-quintillion-bytes-of-data-created-every-day-how-does-cpg-retail-manage-it/}} and CISCO\footnote{\url{https://www.cisco.com/c/en/us/solutions/collateral/service-provider/visual-networking-index-vni/vni-hyperconnectivity-wp.html}} reportedly estimated that significant amount of today's data were created in the past few years. In our current landscape, a significant amount of data is generated and exchanged through mobile and Internet of Things (IoT) devices, such as those found in smart homes and smart cities. There are also corresponding security and privacy risks and challenges associated with such a trend. A typical measure is to perform network traffic monitoring to prioritise and balance traffic load (e.g. to ensure quality of service -- QoS), as well as guaranteeing security and privacy for user data and systems (e.g. by detecting and preventing malicious behaviours).

Currently, network data (data-in-transit) are generally encrypted, as non-encrypted traffic can be subject to a broad range of attacks (e.g., impersonation attack, man-in-the-middle attacks and false data injection attacks), as well as surveillance by both honest-but-curious and malicious threat actors (e.g. violating user anonymity). By analysing the features and characteristics of traffic flows, including encrypted traffic flows, one may be able to identify specific user behaviours \cite{7265055,8006282}. For example, the authors in \cite{Yan2018TrustCom} presented a machine learning-based approach to identify WeChat (a widely used social and payment mobile application) packets and identified transfer transactions by simply analysing the intercepted TLS/SSL traffic. However, current state-of-the-art approaches are generally not capable of differentiating high entropy streams with high accuracy. As later discussed in Section \ref{sec:Background}, only few works focus on distinguishing between encrypted and compressed data, being the work presented in \cite{hahn2018detecting} the most similar to ours as other methods exploit additional features of the collected packages.



Most security and privacy mechanisms rely on (strong) encryption algorithms to protect the communications. However, flawed implementations often use cleartext communication \cite{Wood2017} to transfer sensitive information. The latter is often in IoT devices due to their lack of processing resources. The problem is amplified by the fact that IoT firmware is closed source and cannot be easily extracted since access to the storage and processing units can be concealed or access-protected \cite{yaqoob2019internet}.  The use of compression instead of encryption \cite{hahn2018detecting} from some devices may perplex the problem even more as there are no efficient and accurate methods to distinguish high entropy sources, e.g. encryption from compression. Therefore, a security investigator cannot easily determine whether a device is using a custom compression algorithm or encryption when evaluating its security.

On the contrary, an adversary may penetrate an internal network. To cover his/her actions, the adversary may choose to use compression instead of encryption when exfiltrating data from hosts as encrypted channels may not be available and are not always easy to establish, e.g when pivoting from one machine to another. Note that the compression of data would render many signature-based detection methods useless in network level. The above should also be considered in the context of big data, which may prevent many mechanisms to examine all packages in real-time.

It is, therefore, mandatory to be able to detect encrypted and unencrypted data in network traffic to ensure proper management, auditing and abnormal behaviour detection \cite{Khalife2014,velan2015survey}. Most of state-of-the-art methods are able to detect cleartext/plaintext communications or low entropy file-types (e.g. images and binaries), but are not able to discern between high-entropy files, especially in the case of compressed and encrypted data \cite{Wood2017,hahn2018detecting}. This hinders proper detection of unprotected data being transmitted, which in the case of compressed files can be easily recovered. Therefore, tasks such as detecting devices transmitting unprotected information from individuals, as well as the proper application of privacy policies are difficult to accomplish.

Thus, in this paper we propose a novel traffic classification method named HEDGE, which stands for \textbf{H}igh \textbf{E}ntropy \textbf{D}istin\textbf{G}uish\textbf{E}r. HEDGE is based on the randomness evaluation of the payload information in randomly selected  network traffic packets. Unlike majority of existing approaches, we do not analyse all network traffic. Instead, we only analyse random subsets to enable real-time detection. We select a set of randomness tests from the literature according to their accuracy and efficiency, and implement a threshold-based approach to distinguish between encrypted and non-encrypted high-entropy data streams in real-time.

We also develop a statistically sound dataset using well-known benchmarks and propose a set of strategies to evaluate the utility of our proposal. The dataset has an equally distributed number of file-types; thus, avoiding biases and enhancing the reliability of the outcomes. The code and implementations are publicly available in GitHub\footnote{\url{https://github.com/francasino/traffic_analysis}}.

Based on the evaluation using this dataset, our proposed method correctly classifies 94.72\% of random packets of 64 KB, outperforming current state of the art. Notably, our worst classification results are with small packets of 1 KB each, where the accuracy is slightly above 68.68\%. For instance, the most accurate state-of-the-art work up-to-date that considers random packet inspection \cite{hahn2018detecting} achieves an accuracy of 66.9\% using another sound benchmark, whilst our method achieves an accuracy above 70.6\% using the same dataset. Note that a 4\% is a significant enhancement in this context, especially for small size files, without the use of training or heavy computations. Similar works that study compressed and encrypted packet classification can be found in \cite{6005446} and \cite{Khakpour2013}. More concretely, Lin \cite{6005446} proposes an SVM classifier according to a set of features to identify the type of traffic, including compressed and encrypted files. In the case of encrypted classification, they compute a set of features and train a model using SVM as well as Jensen Shannon divergence (JSD) \cite{61115}. Contrary to our proposal, this approach uses other traffic features to train their model achieving at most 79.8\% accuracy with variable-sized packets, in all cases bigger than 1KB. In \cite{Khakpour2013} authors use a similar approach to the one presented in \cite{6005446} and compare it with a CART decision tree classifier. Nevertheless, they utilise traffic and packet information, as well as the first bits of each packet (i.e. {\em magic header bytes}) where characteristic information of the protocol/method used can be obtained. Although this approach has a series of constraints (e.g. scalability) to be used in real-time classification and thus differs from our approach, authors are able to obtain accuracies near 75\% with SVM and CART and between 76\% and 85\% after testing several feature configurations with different file sizes (their tests consider sizes up to 64KB).

\begin{figure}[th]
    \centering
    \includegraphics[width=\columnwidth]{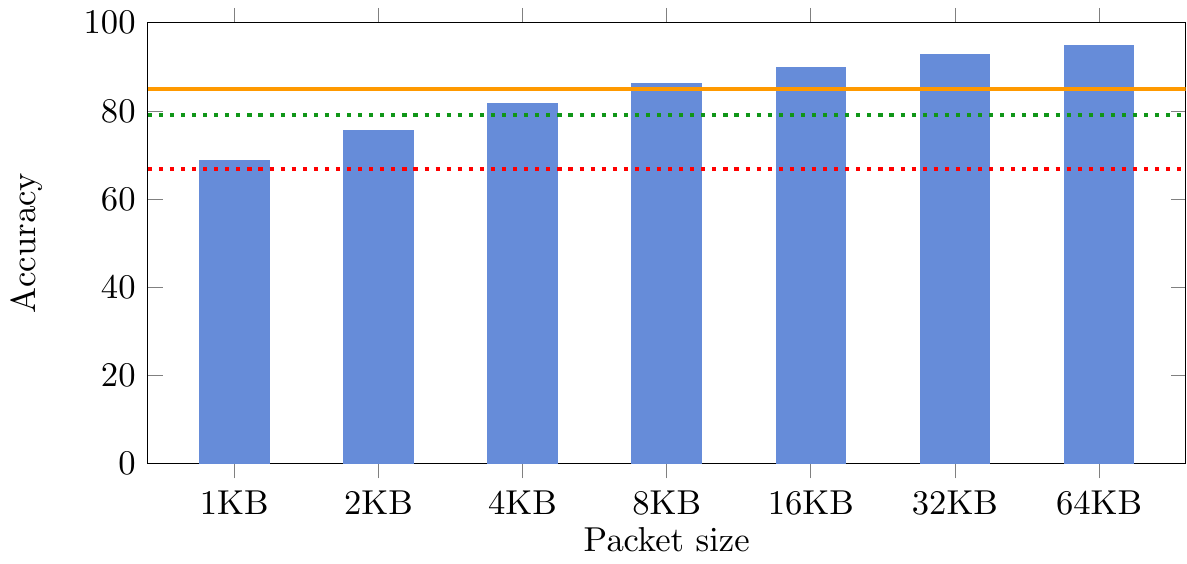}
    \caption{A comparative summary of the classification accuracy of the proposed method for each packet size and that of \cite{hahn2018detecting} (in red), which was only tested for 1KB files. Note that, contrary to us, in the case of \cite{6005446} (in green) and \cite{Khakpour2013} (in yellow), authors use additional traffic information as well as the first {\em magic header bytes} to train their classifiers.}
    \label{fig:comparison}
\end{figure}

A comparative summary of the proposed method with the approaches in \cite{hahn2018detecting}, \cite{6005446} and \cite{Khakpour2013} is presented in Figure \ref{fig:comparison}. Moreover, our method is efficient and easy-to-adopt, avoiding costly training procedures and similarity computations. Further relevant findings when analysing behaviour of compressed and encrypted data depending on the input file-type are also discovered and documented in the paper. For instance, our method manages to distinguish with high accuracy between the file-types from single random packages.

The rest of this paper is organised as follows: Section~\ref{sec:Background} provides the reader with relevant background on network traffic analysis and existing approaches, and widely-used randomness tests. Section~\ref{sec:approach} is devoted to explaining HEDGE, our proposed method. Section~\ref{sec:experiments} describes the experimental setup, the benchmarks and the tests performed to evaluate our proposal. The evaluation findings are later discussed in Section \ref{sec:discussion}. Finally, Section~\ref{sec:conclusions} concludes the paper
and identifies possible directions for future research.


\section{Related Work} \label{sec:Background}

In network traffic analysis approaches, data streams are analysed and classified according to flow and packet information, payload content and statistics (including among others packet size and average times). Based on such information, we can distinguish between: (i) \textit{feature-based} techniques, which determine a set of characteristics/features regarding the traffic flow as well as internal packet information and its structure (i.e. it differs depending on the protocol used), and (ii) \textit{payload-based} techniques, which inspect the data payload content and apply relevant methods to extract statistical properties and identify, for example, the type of information being transmitted (e.g. text, image, video, music, compressed or encrypted). We refer the interested reader to \cite{Khalife2014,velan2015survey} for a classification of network traffic analysis models based on input data, applied techniques applied and their corresponding outcomes.

Most feature-based techniques create a structure or vector containing the features of data flows and then classify them using statistical or machine learning methods \cite{kotsiantis2007supervised,nguyen2008survey}. For instance, in \cite{moore_classification} close to 250 discriminators/features are identified, which can be used for flow record classification.
In the case of payload-based methods, different methodologies to accurately classify data streams according to their payload characteristics \cite{white2013clear} have been studied in the last decade. Machine learning-based techniques generally focus on identifying the application associated with the stream or whether a flow contains some code. Signature-based approaches include hybrid schemes and usually try to exploit characteristics of protocols in some of the first bytes of the packet payloads. In the literature they have been used for identifying widely used application protocols (HTTP, FTP, SMTP) using features such as packet size, port, timing characteristics, and packet payload \cite{early2003behavioral,lee2011netramark,dainotti2009tie,dreger2006dynamic,zhang2000detecting}. Another widely used method is byte frequency distribution, in which each file-type is classified according to its signature \cite{wang2004anomalous,zhang2007approach}. In \cite{Alshammari2009}, the authors evaluated state-of-the art classifiers and compared them with a genetic algorithm. Experiments performed using two benchmarks (SSH and non-SSH samples) show a similar level of accuracy, but genetic algorithm achieves higher efficiency. In \cite{realtimeencrypted}, the authors used $k$-nearest neighbours ($k$-NN) and $k$-means for flow classification. Their implementation enables efficient similarity computation, since only $c$ comparisons are required (considering that files have been classified in $c$ clusters) to find the closest flows and classify them accordingly.

Entropy-based classification methods have also been the focus of ongoing research \cite{tabish2009malware,weber2002toolkit,stolfo2005fileprint,lyda2007using}. In addition to the use of Shannon entropy \cite{shannon1949communication} or the chi-square test \cite{d2017goodness} to classify data streams, approaches such as those reported in \cite{roussev2009file,haggerty2007forsigs,li2006forensic} have been used to model the structure of files. Such techniques, however, incur high computational costs for outcome validation. The costs also increase with the amount of traffic analysed.

Although encrypted traffic hides all content from a potential eavesdropper, the first steps (initial handshake, connection parameter negotiation, protocol version) are performed in plaintext and therefore have distinguishable features. Exploiting the latter, machine learning and other classification methods can infer the underlying application protocols from encrypted traffic \cite{alshammari2011can,dorfinger2011entropy,wright2006inferring,5996087}, for example in SSH and web browsing traffic analysis \cite{liberatore2006inferring,sun2002statistical}. Specifically, classification algorithms used to detect encrypted web traffic rely on static object lengths and the previous creation feature-based libraries (i.e. training models). An associated limitation is that these methods  are not reliable for dynamic encryption schemes applied to different file sizes and lengths, padding or other techniques. In the work of Zhang et al. \cite{zhang2007analyzing}, the authors analysed browsing traffic to detect self-decrypting exploit code. In \cite{5945941}, the authors used three techniques (i.e. a multi-objective genetic algorithm (MOGA) \cite{murata1995moga}, the C4.5 algorithm \cite{ruggieri2002efficient} and a semi-supervised k-means \cite{Erman2007}) on different benchmark datasets to classify encrypted data streams on SSH. They performed binary identification of SSH traffic in the following two settings: (i) using test data from the training model, and (ii) using test data from a different network. Findings from their evaluation showed that MOGA performs better in the first case, whilst C4.5 outperforms the rest in the second. In \cite{6785319}, the authors characterised the differences between encrypted and non-encrypted traffic and designed a function to correlate the encrypted and non-encrypted traffic according to their features, as well as the minimum number of packets to guarantee an accurate classification.

Encrypted data is known to be more uniformly distributed than unencrypted data \cite{malhotra2007detection,shannon1949communication}. Therefore, a number of approaches rely on this characteristic to locate cryptographic keys stored in memory and file system dumps \cite{shamirplay}. The authors generally analysed big data streams, divided them into small blocks, and computed their entropy. Therefore, high entropy blocks may indicate the presence of encrypted data. Similarly, the authors in \cite{olivain2006detecting} and \cite{dorfinger2010entropy} first computed the entropy of packet payloads and then compared it with the entropy of uniformly randomly distributed sequences of the same length. However, the entropy estimation approach is not effective when the number of samples is small \cite{paninski2003estimation,paninski2004estimating}. Moreover, entropy measures are not reliable when other data which has high entropy is present such as compressed, MP3 or PDF files.

There has been renewed interest in distinguishing between compressed and encrypted data streams, partly due to the increasing use of end-to-end encryption in online communications. As discussed earlier, one of the most prevalent strategies is to use Shannon entropy or the chi-square test, among other measures over fixed-size sliding data streams (1KB, 2KB and so on) to distinguish between different types of data, as well as compressed and encrypted. However, one of the most challenging aspects, beyond distinguishing between high entropy files, is the time required to develop this classification in continuous traffic monitoring; thus, limiting its applicability in real-time scenarios.

As claimed by Malhotra et. al. \cite{malhotra2007detection}, most traffic analysis tools analyse the contents of the data packets but they are not capable of determining whether data are encrypted. According to their research findings, Shannon entropy can be used to accurately classify low entropy streams. However, other techniques such as the chi-square test are needed to differentiate high entropy streams. Moreover, the authors noted that many compression algorithms have a fingerprint as their first bytes ({\em magic header bytes}) imply their format and thus, such information can enhance classification's accuracy. In \cite{6005446}, the authors used entropy and frequencies of characters to distinguish the type of payload content (e.g. text, picture, audio, video, compressed, base64-encoded image, base64-encoded text and encrypted), with small computing space and notable accuracy. In \cite{Khakpour2013}, the authors claimed that if the probability distribution of each subset of a file is similar, we can distinguish such file from others. However, such differences may be too small in the case of high-entropy compressed files and encrypted files, especially if we consider small files ($<$ 4KB). In another work \cite{conti2010automated}, the authors performed $k$-NN classification to distinguish between random, compressed and encrypted data streams. In \cite{182951}, authors use chi-square test to distinguish between encrypted and encoded (compressed) data to bypass DRM
protection in streaming media services. Nevertheless, authors need to use a considerable amount of data (i.e. more than 380KB) to ensure reliable classification outcomes.

Existing methods generally rely on continuous traffic information to enhance their accuracy by collecting information about \textit{complete} packet transmission, the beginning and the end of a connection or of a file and etc. Therefore, real-time monitoring is inefficient using such schemes, since they require the analysis of huge volumes of data. However, these schemes can be useful in studying past-events or analysing only specific connections. Hence, to enable real-time monitoring, our aim is to analyse the payload of a random subset of packets and infer as much information as possible. Similar to our approach, the work of Hahn et al. \cite{hahn2018detecting} presents the first, to the best of our knowledge, technique to distinguish encrypted from compressed unencrypted network transmissions by analysing random packets. The authors applied three machine learning models, with the convolutional neural network (CNN) model achieving the best results. However, more efficient solutions are needed to perform practical real-time traffic classification even in the case of random data stream analysis, as the methods applied in \cite{hahn2018detecting} (i.e. CNN and $k$-NN) are computationally expensive and require proper training.

\section{Proposed Approach}
\label{sec:approach}
In this section, first we state the assumptions that we make for our approach e.g. capabilities of the entities involved and the requirements that are needed from any such a solution. Then, we discuss how we perform the feature selection and the steps of our methodology.

\subsection{Basic assumptions and requirements}

\begin{figure}[th]
    \centering
    \includegraphics[width=\columnwidth]{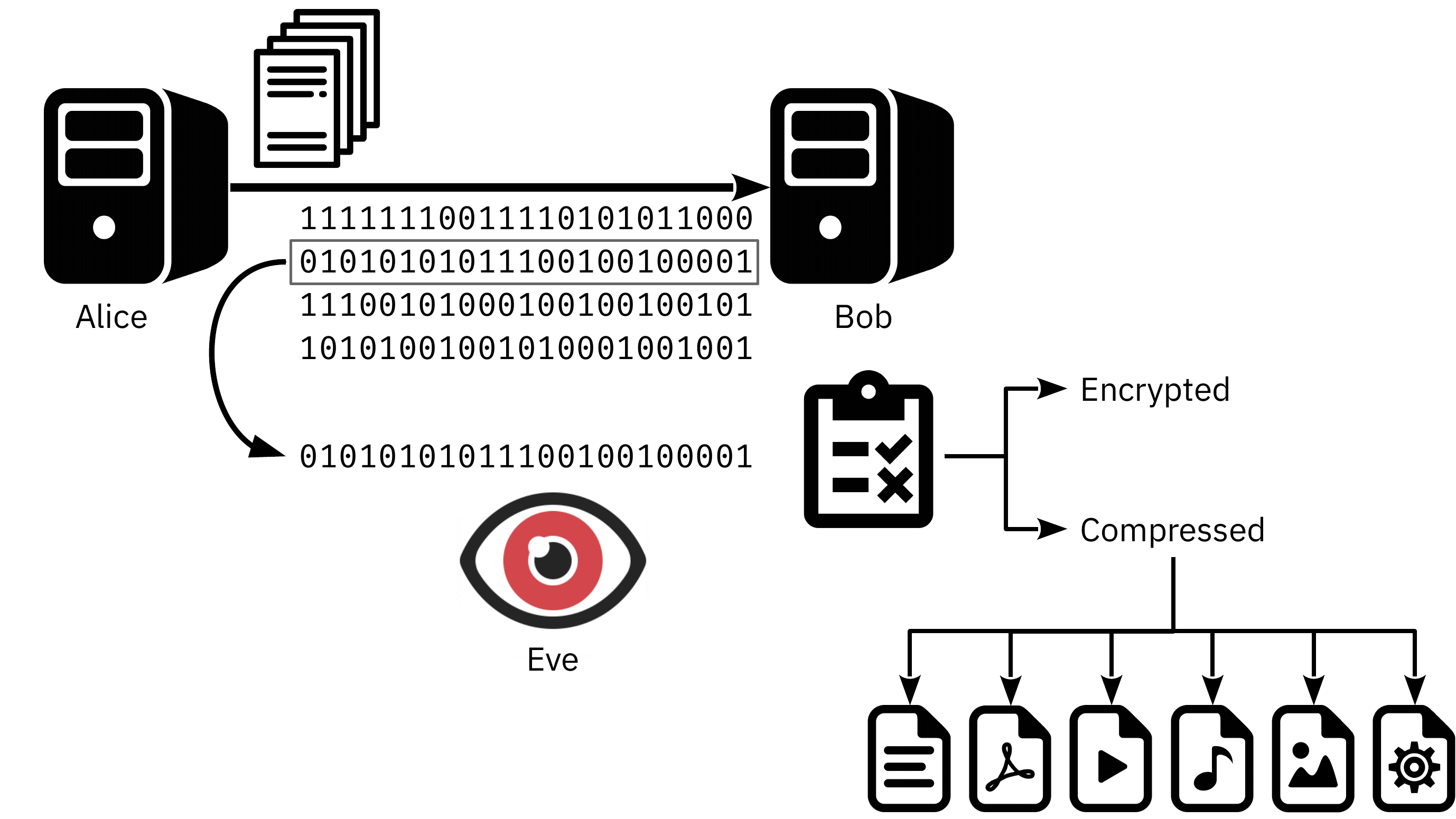}
    \caption{An overview of the proposed approach.}
    \label{fig:classifier}
\end{figure}

In our model, we assume  \textit{Alice} and \textit{Bob} are exchanging messages over a public channel and the exchanged message can either be plaintext, encrypted or compressed. Therefore, messages where one part is encrypted and the other part plaintext are beyond the scope of this study. Moreover, we assume that another entity, \textit{Eve} (i.e. the adversary), can passively intercept the traffic between Alice and Bob. Typically in the literature, Eve is considered malicious, however, we do not assume malicious motives for her actions. That is because Eve might be a firewall, intercepting the traffic only to determine its content and block malicious requests or packets that may expose the nodes of the internal network to threats. Additionally, we assume that Eve cannot determine whether two or more packets belong in the same message or session, therefore, each packet is studied \emph{individually}. The latter enables a more generic scenario with a weak assumption, compared to the bulk of the literature \cite{6005446,Khakpour2013}. Therefore, we assume that Eve can analyse only a fragment of packets due to e.g. high volume of network traffic or processing constraints. Figure \ref{fig:classifier} shows an overview of the proposed scenario.

Furthermore, in order to enhance the applicability of our solution, we require the following properties:
\begin{itemize}
    \item \textbf{Accurate:} The proposed solution should be able to distinguish between plaintext, compressed and encrypted files with high accuracy. If possible, the solution should be able to differentiate file-types (image, binary, etc.) as well.

   \item \textbf{Efficient:} The performed tests or methods must provide fast and robust responses. In addition to method efficiency, we want to implement a tree-based threshold system, so that in most cases, we do not need to run all tests to discard or classify an element (message); thus, improving the efficiency of the method.

   \item \textbf{Adaptable:} The proposed  solution must allow fine-tuning of parameters, depending on the network traffic, to achieve better classification accuracy than static methods.

   \item \textbf{Reproducible:} The proposed methodology has to be easy to implement and reproduce, in order to facilitate adoption and verification. To address this, we will use state-of-the-art methods and a set of predefined strategies to automate the database creation and randomness tests.
\end{itemize}

\subsection{Feature Selection}
\label{sec:featureselection}
After evaluating the performance and the computational cost of the methods described in Appendix \ref{sec:randomnesstests} using random data from well-known benchmark datasets (later described in Section \ref{sec:experiments}), only a subset of them met the requirements to be used in HEDGE. More precisely, the chi-square test can efficiently recognise true random bit streams from non-random ones and the \% of confidence is a reliable indicator, exhibiting (according to our architecture) a better trade-off between computational time and accuracy than other distribution-based approaches such as Kolmogorov-Smirnov and Anderson-Darlong. FIPS-2-140 tests are very stable in the case of encrypted bit streams, that is passing the FIPS-2-140 test without fails. After discarding the common tests of SP 800-22 and FIPS-2-140, as they were already performed in the latter, we selected a small subset of SP 800-22 tests that could efficiently distinguish between compressed and encrypted files, without the 20.000 bits limitation though to accommodate for small packets. After experimenting with all tests of this suite, we identified that the ones that fit the requirements were the frequency within block test, the cumulative sums test, and the approximate entropy test.

The average and correlation tests, although having slightly more accuracy and less variability in encrypted than in compressed streams, do not provide reliable outcomes for small size files with high entropy. The other remaining methods could not be used as either (1) they did not provide a good indicator -- both encrypted and compressed files had indistinguishable results, considering the standard deviation of the outcomes-- or (2) they are computationally costly. For instance, the entropy is a good indicator only between raw files and the rest, according to our tests and \cite{Wood2017,hahn2018detecting}, and the Diehard most common tests (e.g. birthday spacing, parking lot or random spheres) are usually passed by both compressed and encrypted files. Moreover, Diehard tests, as well as TestU01 tests, require a considerable amount of time to be computed and thus, they are not suitable for real-time traffic analysis. A summary of the selected features is described in Table \ref{tab:features}.

\begin{table}[tb!]
\small
\renewcommand{\arraystretch}{1.25}
\renewcommand{\tabcolsep}{1mm}
  \centering
   \caption{Features selected from each randomness test. Therefore, a total of 3 features will be used to define the threshold. Note that we aggregate the outcomes of NIST tests.}
  \begin{tabular}{ll} %
    \toprule
 \multicolumn{1}{c}{\textbf{Method}}   &
 \multicolumn{1}{c}{\textbf{Features}}  \\
    \midrule
\textbf{Chi square test}   & Absolute value \\
\textbf{Chi square test}   & $\chi\%$ of confidence \\
\textbf{NIST SP 800-22}   &  Aggregate number of failed blocks\\
    \bottomrule
  \end{tabular}

  \label{tab:features}%
\end{table}

\begin{table}[tb!]
\small
\renewcommand{\arraystretch}{1.25}
\renewcommand{\tabcolsep}{0.5mm}
  \centering
   \caption{Threshold values for each selected feature. The value $x$ denotes the outcome of the randomness test for a single bit stream.}
  \begin{tabular}{c|c|c} %
    \hline
 \multicolumn{1}{c}{\textbf{Chi abs. val.}}   &  \multicolumn{1}{c}{\textbf{$\chi\%$}}   & \multicolumn{1}{c}{\textbf{NIST SP 800-22}}  \\
 \hline
   $x \in AVG \pm \sigma$  & $\chi\%>99\%$ $\|$ $\chi\%<1\%$ & 0 fails \\
    \hline
  \end{tabular}

  \label{tab:thresholdvalues}%
\end{table}

To enable data payload classification, HEDGE implements a threshold-based method that takes into account the set of features selected from Table \ref{tab:features}. The threshold values selected for each feature are described in Table \ref{tab:thresholdvalues}. More concretely, we consider the average and its standard deviation $\sigma$ for the chi square test. The $\chi\%$ most relaxed confidence (i.e. $\chi\%>99\%$ $\|$ $\chi\%<1\%$) is also considered (see Section \ref{sec:randomnesstests}). Finally, we set the threshold of the number of NIST SP 800-22 tests failed to 0. The values for each category have been selected after analysing the outcomes of a huge set of random encrypted files for each test. The threshold values selected reflect the stability of their outcomes regardless of the input file-type (e.g. text, image, binary and etc). Therefore, the results of the tests are almost indistinguishable between encrypted files, a feature which is discussed with more details in Section \ref{sec:experiments}. 

\begin{figure*}[th!]
\centering
\subfloat[\% of accuracy for encrypted detection in 64KB files.		\label{fig:64kbencryp}]{\includegraphics[width=.4\textwidth]{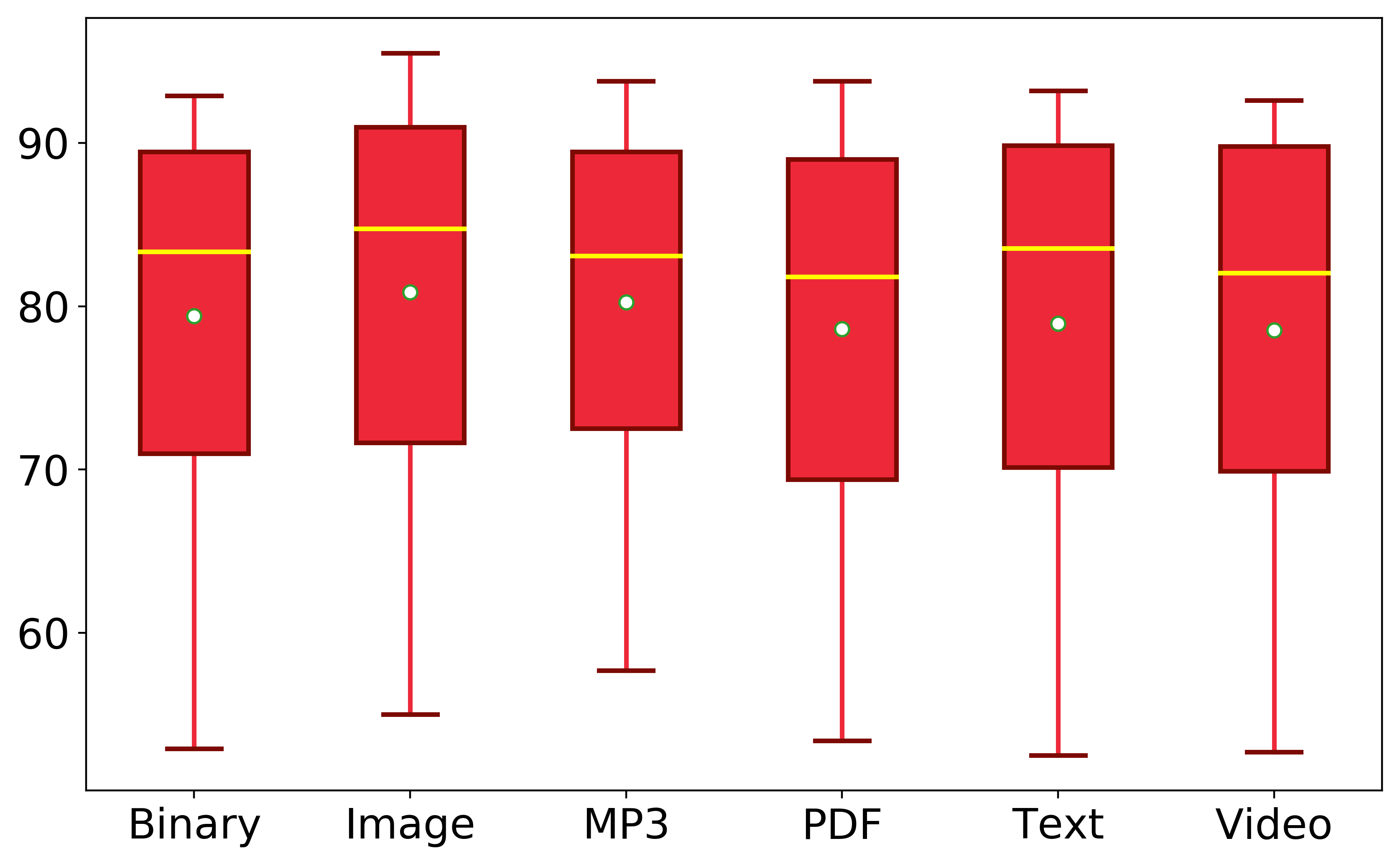}}
\subfloat[\% of accuracy for compressed detection in 64KB files.  		\label{fig:64kbcomp}]{\includegraphics[width=.4\textwidth]{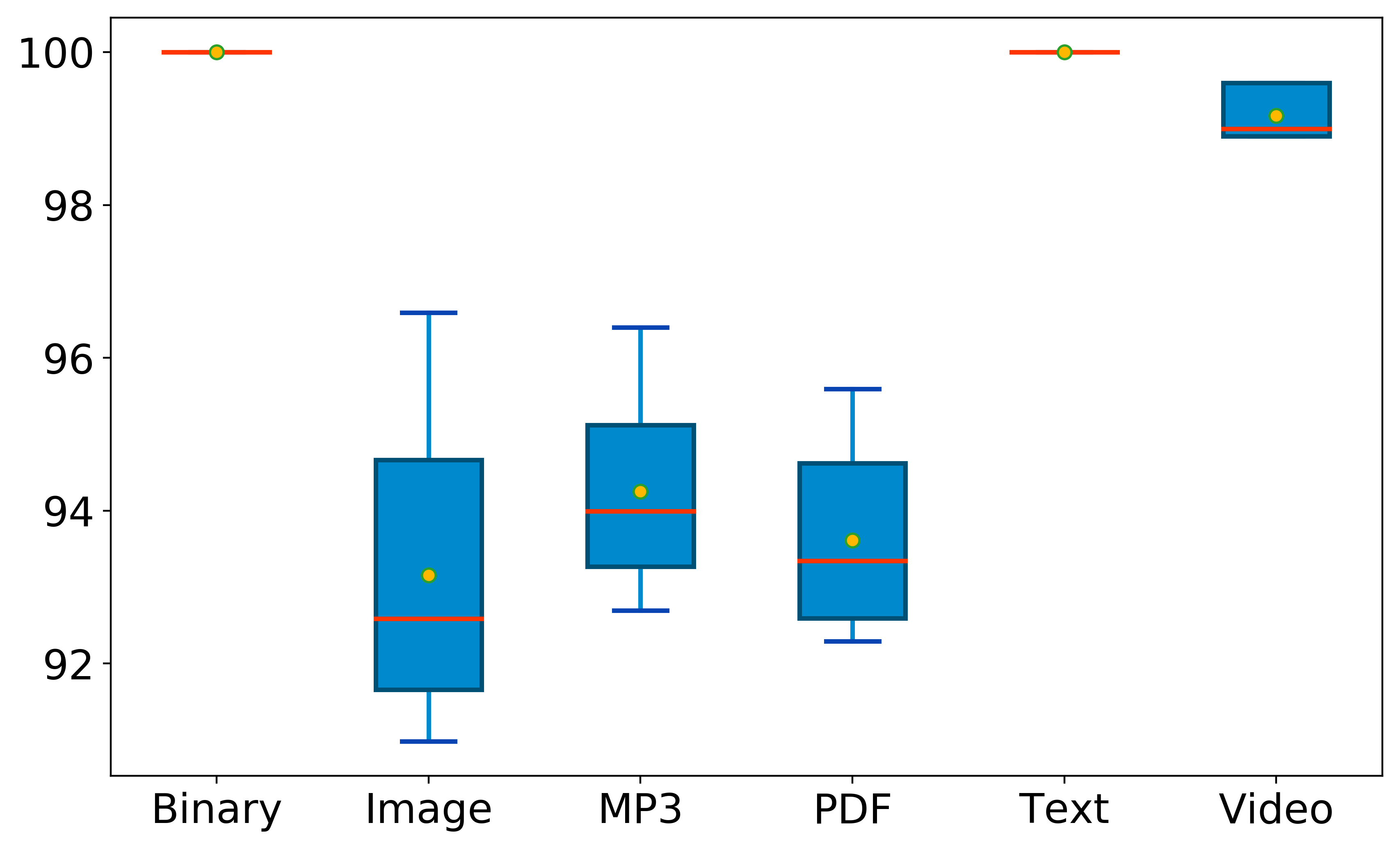}}\\[-2ex]
\subfloat[\% of accuracy for compressed detection in 32KB files.  		\label{fig:32kbcomp}]{\includegraphics[width=.4\textwidth]{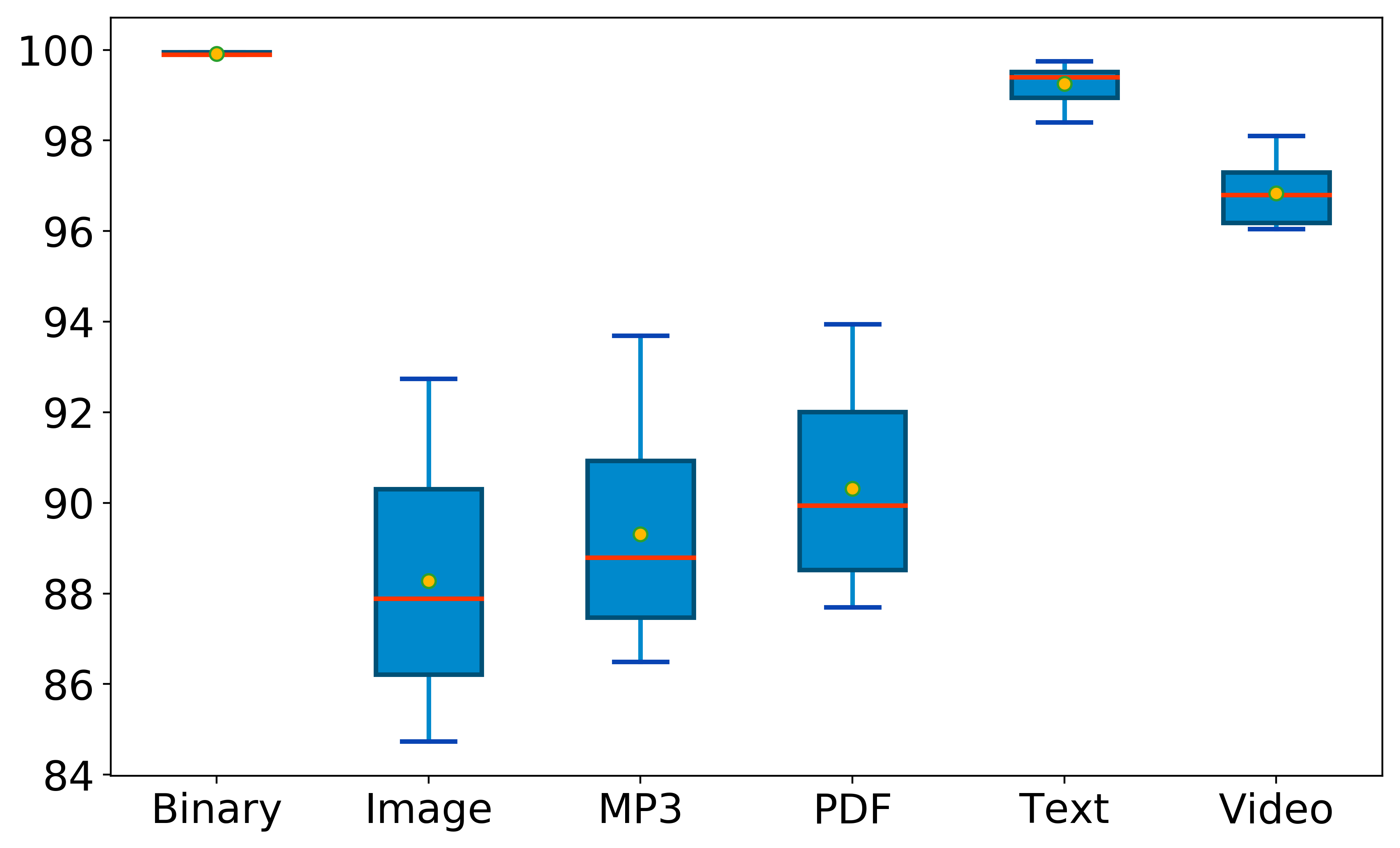}}
\subfloat[\% of accuracy for compressed detection in 16KB files. 	\label{fig:16kbcomp}]{\includegraphics[width=.4\textwidth]{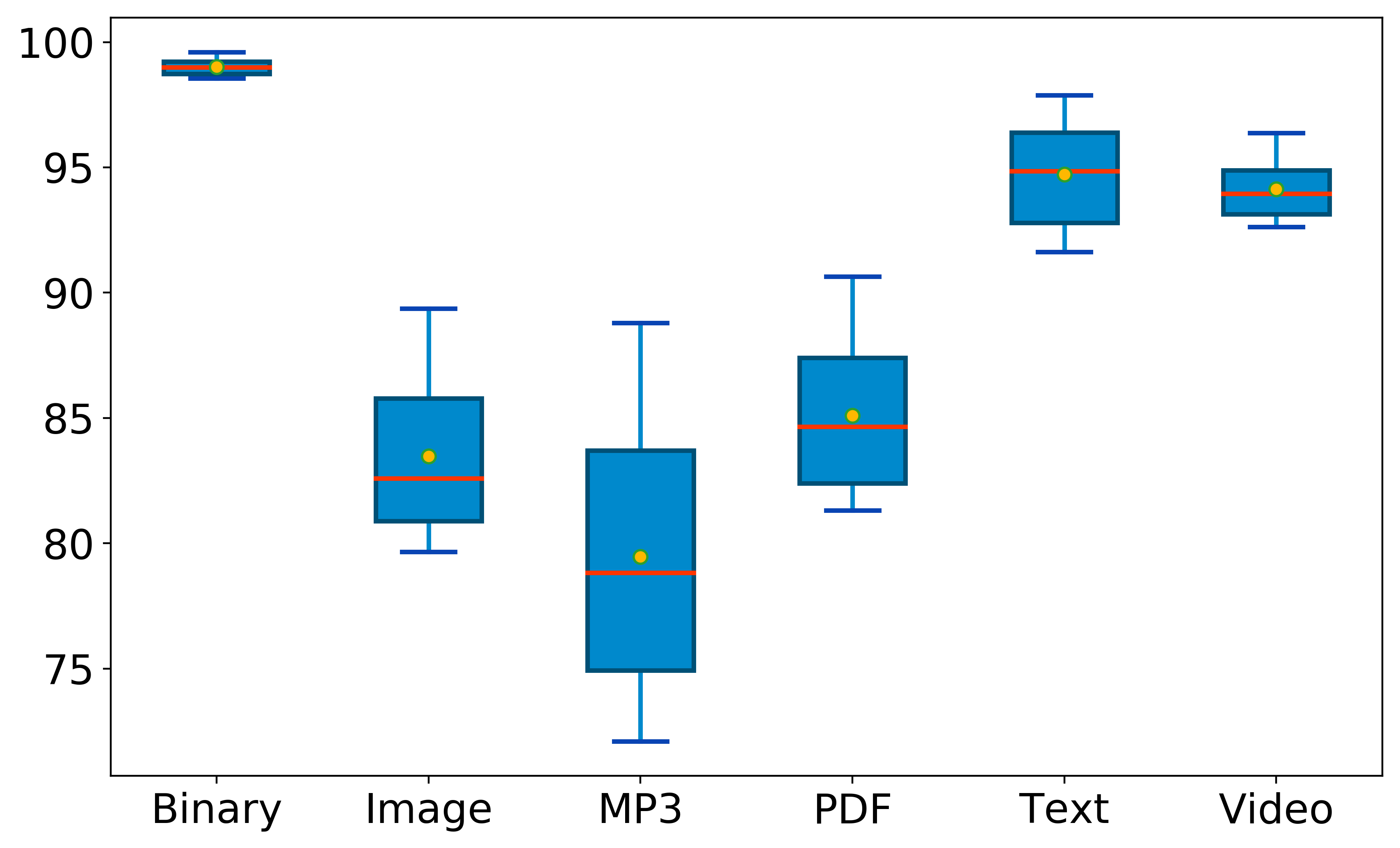}}\\[-2ex]
\subfloat[\% of accuracy for compressed detection in 8KB files.  		\label{fig:8kbcomp}]{\includegraphics[width=.4\textwidth]{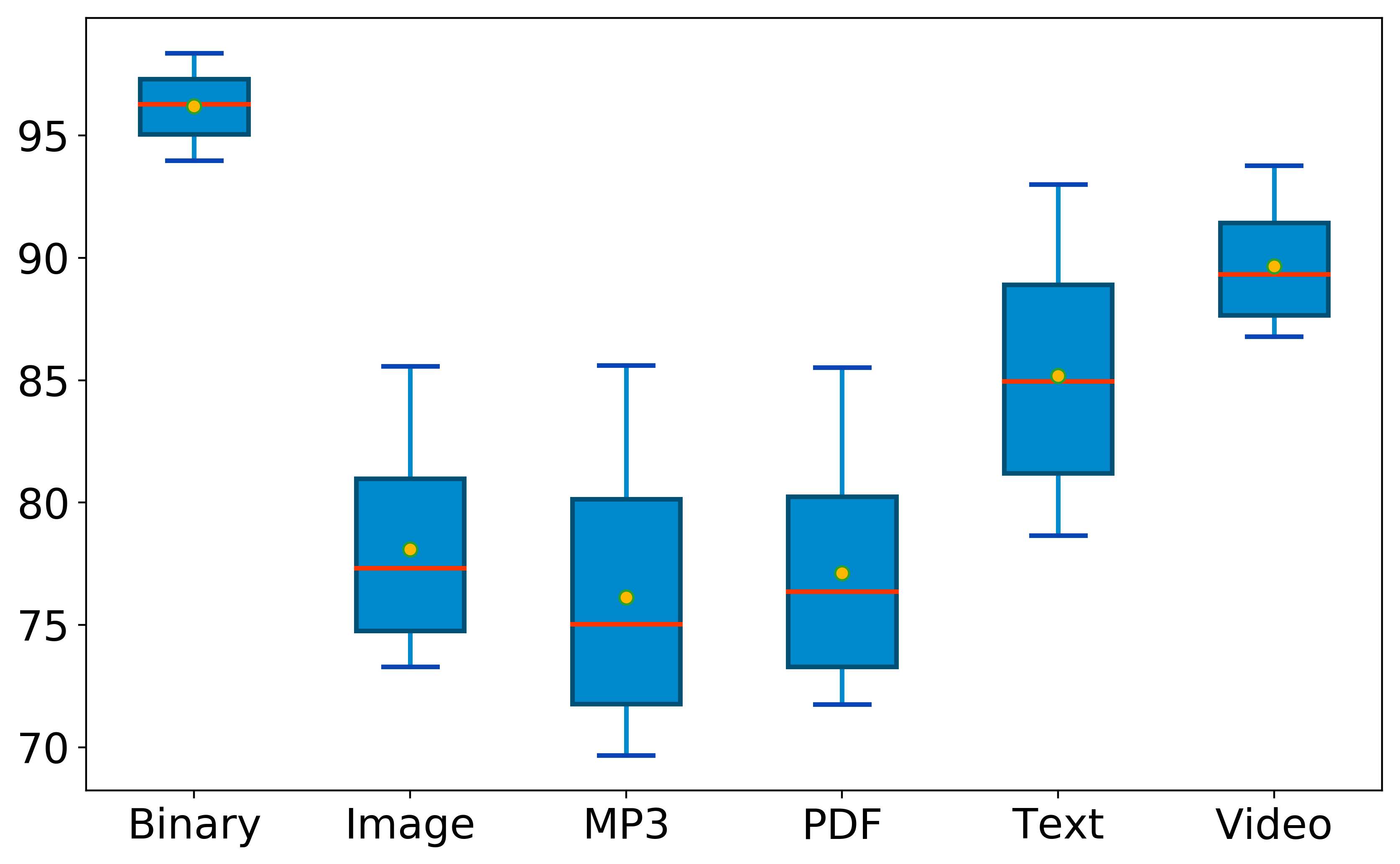}}
\subfloat[\% of accuracy for compressed detection in 4KB files. 	\label{fig:4kbcomp}]{\includegraphics[width=.4\textwidth]{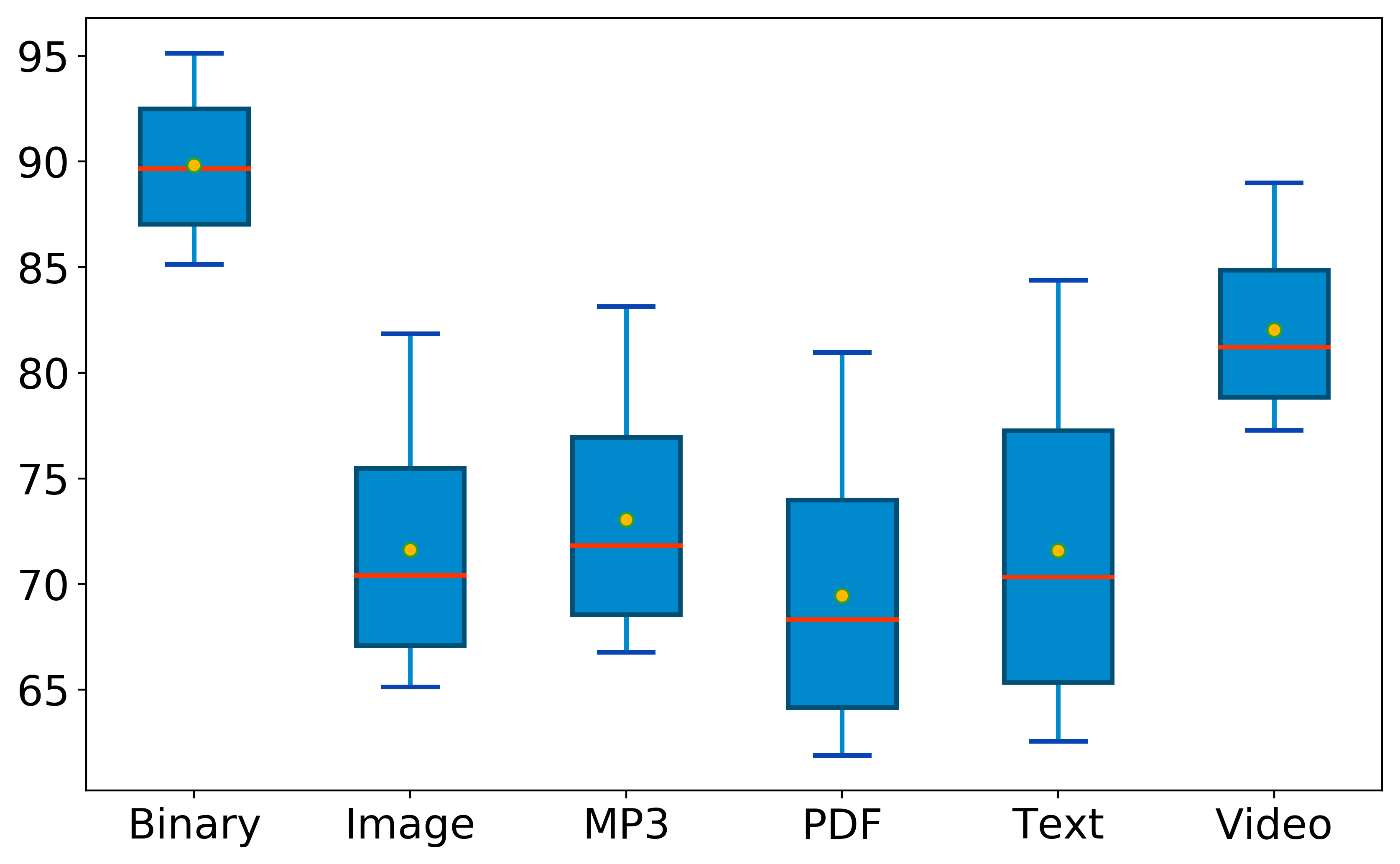}}\\[-2ex]

\subfloat[\% of accuracy for compressed detection in 2KB files.  		\label{fig:2kbcomp}]{\includegraphics[width=.4\textwidth]{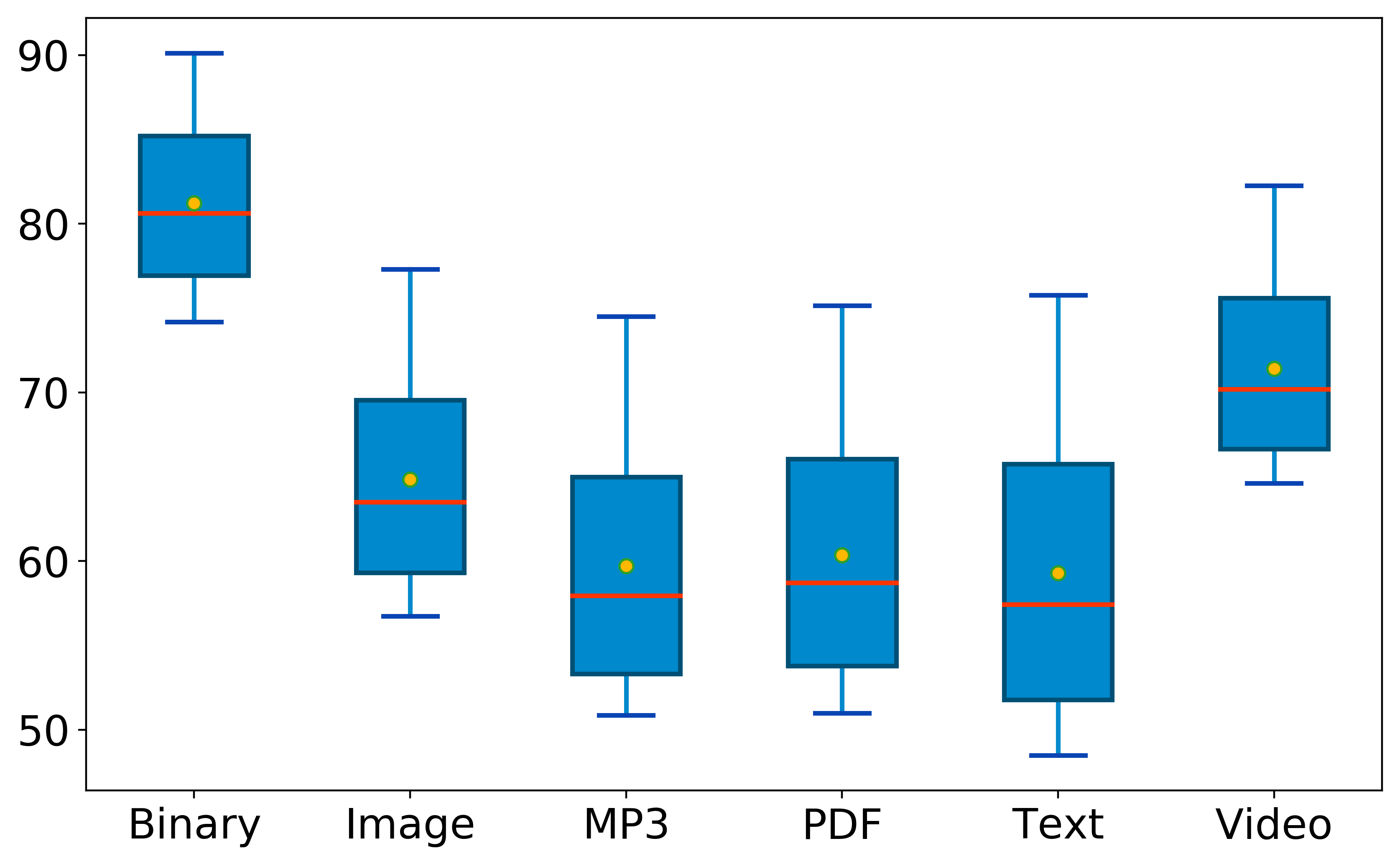}}
\subfloat[\% of accuracy for compressed detection in 1KB files. 	\label{fig:1kbcomp}]{\includegraphics[width=.4\textwidth]{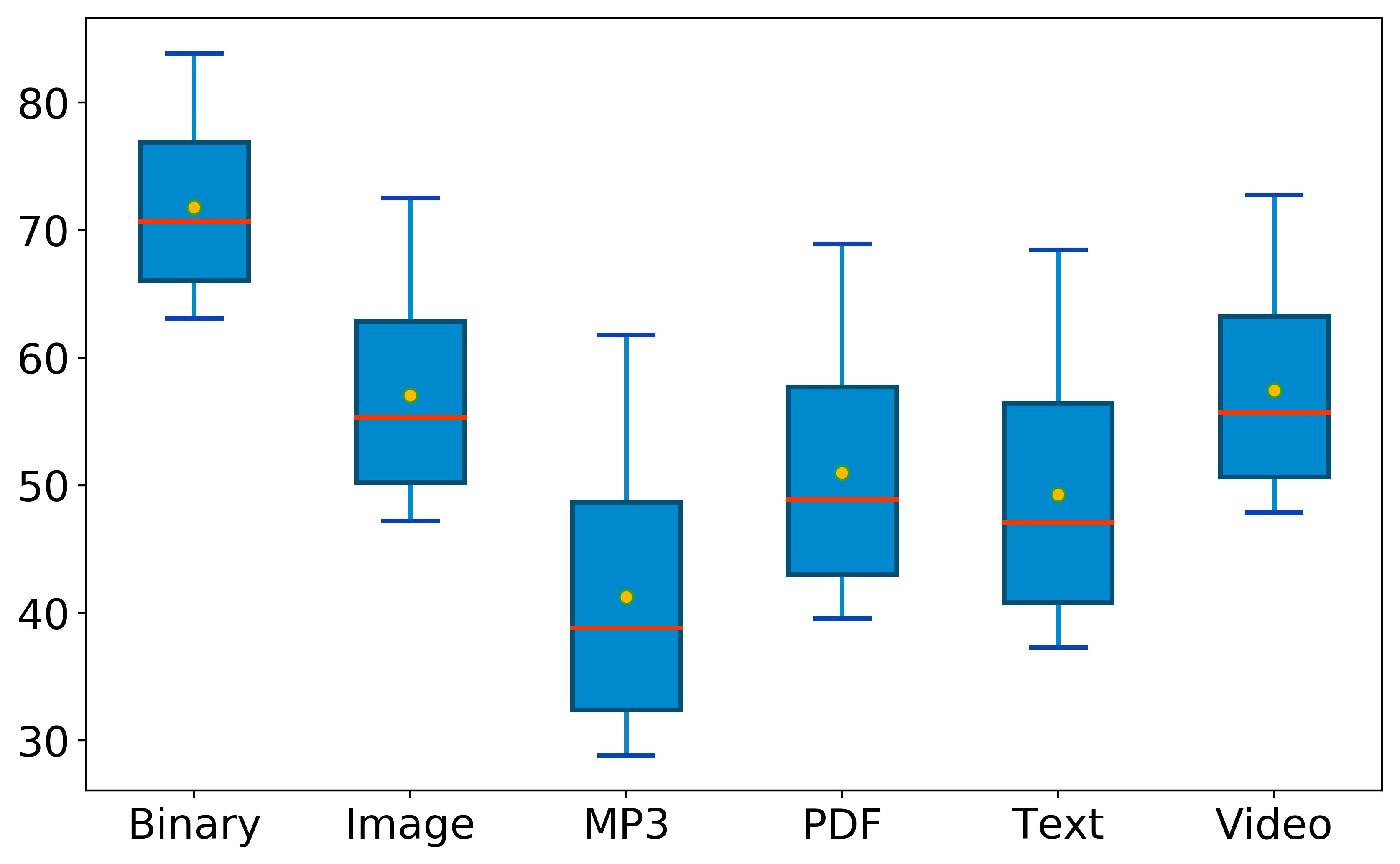}}

\caption{Detail of the accuracy according to input file-type in compressed and encrypted files. Lines represent the median whilst points represent the average value.}
\label{fig:filetypedetail}
\end{figure*}

To introduce some variability and provide a more adaptable threshold to the input bit streams, HEDGE adds a {\em gain} factor $\gamma$ to the threshold values. Therefore, we can apply relaxed or strict policies in our detection system (as later described in Figure \ref{fig:filetypedetail} and Table \ref{tab:detailedoutcomes}), observe its behaviour and select the one that provides better outcomes according to the input data. In this regard, $\gamma$ is added as a factor of the standard deviation of the threshold values in the case of the chi-square absolute value. The remaining features will not be affected by $\gamma$. For instance, the chi-square \% confidence should be a fixed factor, as well as the NIST SP 800-22 test fails. Note that accepting one fail in NIST SP 800-22 test implies a considerable amount of false positives.

\begin{figure}[tb]
	\begin{center}
		\includegraphics[width=1\columnwidth]{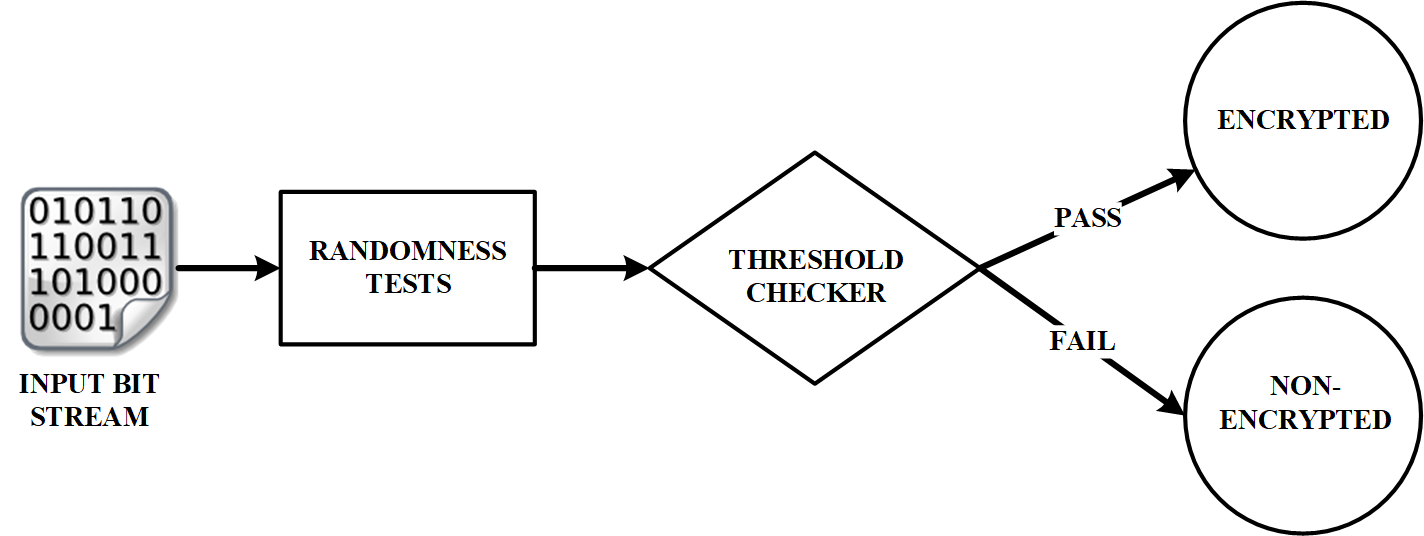}
		\caption{Step by step of our methodology.}
		\label{fig:overview}
	\end{center}
\end{figure}

An overview of our method is depicted in Figure \ref{fig:overview}. First, we apply the randomness tests to the input bit stream and check whether the outcomes pass our threshold (according to a desired level and $\gamma$). We consider the data stream encrypted if, and only if, it passes all thresholds. Note that our approach can be easily parallelised, since it can be executed in independent machines regardless of the input data. Therefore, there is no need for a centralised database to compute distances/similarities, such as in the case of $k$-NN or clustering methods.

\section{Experiments}
\label{sec:experiments}

In this section, we detail the experimental setup and the tests performed to validate the efficacy of HEDGE. In Section \ref{sec:discussion}, we present the results.

As stated in Table \ref{tab:features}, we use the chi-square and a subset of NIST SP 800-22 tests. However, since the outcomes of some tests vary depending on the file size, we apply our method to different bit stream sizes.

The first step of the experiment is to create a statistically sound dataset. Therefore, we use a random set of files retrieved from well-known benchmarks, summarised in Table \ref{tab:datasets}, to create a heterogeneous dataset. Our dataset has the same number of files for each file-type (i.e. image, video, binary, audio, text and PDF) to avoid biases, since the randomness of compressed files is input-dependant (see Section \ref{sec:discussion}). Note that we selected the most representative file-type according to the state-of-the-art \cite{white2013clear,hahn2018detecting,Khakpour2013}.

\setlength{\textfloatsep}{3pt}
\begin{table}[tb!]
\small
\renewcommand{\arraystretch}{1.25}
\renewcommand{\tabcolsep}{1mm}
  \centering
   \caption{Set of encryption and compression methods. Considering all encryption and compression variants, and each input generates a total of 10 different new files.}
  \begin{tabular}{c|c} %
    \hline

 \multicolumn{1}{c}{\textbf{Encryption}}   & \multicolumn{1}{c}{\textbf{Compression}} \\
    \hline
\multirow{2}*{}  \textbf{AES}   (128 / 192 / 256) & \textbf{ZIP}   \textbf{RAR}   \textbf{BZIP2}   \textbf{GZIP} \\
  \textbf{Camelia}    (128 / 192 / 256)   & \\
\hline
  \end{tabular}

  \label{tab:compencmethods}%
\end{table}



\setlength{\textfloatsep}{3pt}
\begin{table}[tb!]
\small
\renewcommand{\arraystretch}{1.25}
\renewcommand{\tabcolsep}{1mm}
  \centering
   \caption{Datasets used to obtain raw files for our experiments. All data have been randomly selected. }
  \begin{tabular}{ll} %
    \hline
 \multicolumn{1}{c}{\textbf{File-type}}   &
 \multicolumn{1}{c}{\textbf{Benchmark}}  \\
    \hline
\textbf{IMG}   & \textbf{COCO Dataset \footnote{\url{http://cocodataset.org/home}} }   \\
\textbf{IMG}   & \textbf{Microsoft Research \footnote{\url{https://www.microsoft.com/en-us/research/project/rgb-d-dataset-7-scenes/}}}    \\
\textbf{PDF}   & \textbf{ArXiv \footnote{\url{https://arxiv.org/}}}    \\
\textbf{TXT}   & \textbf{Project Gutenberg \footnote{\url{https://www.gutenberg.org/}}}    \\
\textbf{MP3}   & \textbf{9th Symphony of Beethoven}  \\
\textbf{VIDEO}   & \textbf{Cambridge-driving Labeled Video Database \footnote{\url{http://mi.eng.cam.ac.uk/research/projects/VideoRec/CamVid/}}}    \\
\multirow{2}*{\textbf{BIN/EXEC}}      & \textbf{/System32 in Win10 x64}\\
& \textbf{/sbin in Ubuntu 16.04}   \\

    \hline
  \end{tabular}

  \label{tab:datasets}%
\end{table}

Next, since the aim of the paper is to classify high entropy files, we generate a set of compressed and encrypted bit streams of fixed size which covers all powers of 2, starting from 1KB and up to 64KB. To cover the variability of real-life scenarios, we use a set of prevalent encryption and compression methods, which are summarised in Table \ref{tab:compencmethods}. The dataset creation step is described in Algorithm \ref{algorithm}. Both arrays $Sizes$ and $Methods$ used in Algorithm \ref{algorithm} {\em line 1} correspond to the different payload sizes (from 1KB to 64KB) and to the contents of Table \ref{tab:compencmethods}, respectively. After applying Algorithm \ref{algorithm} to the data extracted from the benchmarks described in Table \ref{tab:datasets}, the result is a collection of datasets (one for each file size), where each contains approximately 1GB of compressed and encrypted files. To further guarantee equally distributed amount of files and unbiased experiments, we randomly select a subset with the same number of encrypted and compressed files to perform our experiments. Therefore, our datasets consist of exactly 50\% encrypted and 50\% compressed files.

\begin{algorithm}[!tb]
\begin{algorithmic}[1]
\caption{Database Creation}
 \Function{Generate Dataset}{ DataSet D, Array Sizes,
 Array Methods}\Comment{The possible bit stream sizes and the compression and encryption methods} \\
\While{(DatasetHasFiles)}
 \State $f_i$ = SelectTheNextFile (D);\Comment{Next raw file in dataset}
 \State $V$ = CreateFileVariants ($f_i$, Array Methods); \Comment{Creates a set of files for each $f_i$}
 \State $S$ = SplitFiles ($V$, Array Sizes); \Comment{Each file in V is split into different sizes}
 \EndWhile
\EndFunction
\label{algorithm}
\end{algorithmic}
\end{algorithm}

\setlength{\textfloatsep}{3pt}
\begin{table}[tb]
\small
\renewcommand{\arraystretch}{1.35}
\renewcommand{\tabcolsep}{1mm}
  \centering
   \caption{Threshold values for each selected feature. Values correspond to the values obtained considering all training sets of our experiments.}
  \begin{tabular}{c|c|c|c} %
    \hline
 \multicolumn{1}{c}{\textbf{KB}} & \multicolumn{1}{c}{\textbf{Chi-square abs. val.}}   &  \multicolumn{1}{c}{\textbf{$\chi\%$}}   & \multicolumn{1}{c}{\textbf{NIST SP 800-22}}     \\
 \hline
  64 &  $255.37 \pm 22.82$  & $x>99$ $\|$ $x<1$ & 0 fails \\
  32 &  $255.08 \pm 22.68$  & $x>99$ $\|$ $x<1$ & 0 fails   \\
   16 & $254.96 \pm 22.76$  & $x>99$ $\|$ $x<1$ & 0 fails \\
   8 &  $255.09  \pm 22.54$  & $x>99$ $\|$ $x<1$ & 0 fails  \\
  4 &  $255.04  \pm 22.60 $ & $x>99$ $\|$ $x<1$ & 0 fails\\
  2 & $254.98  \pm 22.57$  & $x>99$ $\|$ $x<1$ & 0 fails  \\
  1 &  $255.02  \pm 22.57$  & $x>99$ $\|$ $x<1$ & 0 fails   \\

    \hline
  \end{tabular}

  \label{tab:thresholdpersize}%
\end{table}


\begin{figure}[th!]
\centering
{\includegraphics[width=\columnwidth]{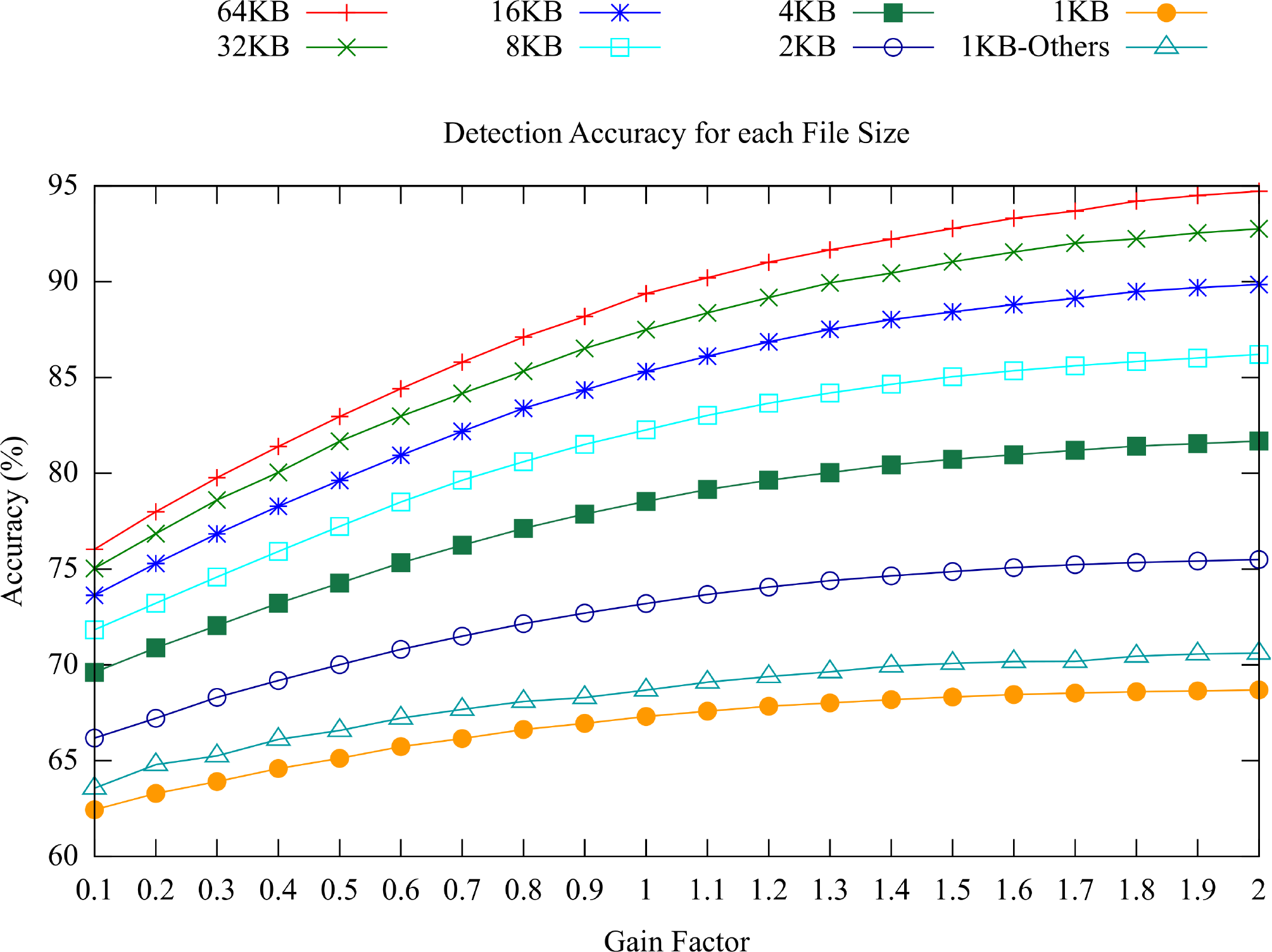}}
\caption{Accuracy detection according to the different file sizes and the parameters of our system.}
\label{fig:accuracy}
\end{figure}

Next, we use an inverse 10-fold experiment on each dataset and repeat the experiments 1000 times. More concretely, a 10\% of data is randomly selected (5\% of compressed and 5\% of encrypted files) and the threshold levels are trained using the encrypted data, while the 5\% of compressed data is discarded. Hence, the remaining data (90\%) is used for testing and validation. The possible outcomes of our classification are summarised in Table \ref{tab:possibleoutcomes}. Therefore, the accuracy of our method will be computed as the sum of $TP+TN$. In addition to our benchmark, we compute the accuracy of our method using another state-of-the-art dataset \cite{hahn2018detecting}.

\begin{table}[tb!]
\small
\renewcommand{\arraystretch}{1.25}
\renewcommand{\tabcolsep}{1mm}
  \centering
   \caption{Possible outcomes of our method according to the input file-type.}
  \begin{tabular}{rcl} %
    \toprule
 \multicolumn{1}{c}{\textbf{Input}}   &
 \multicolumn{1}{c}{\textbf{Output}}  & \multicolumn{1}{c}{\textbf{Result}} \\
    \midrule
\textbf{Encrypted File}   & PASS & true positive (TP) \\
\textbf{Encrypted File}   & FAIL & false negative (FN) \\
\textbf{Compressed File}   & PASS & false positive (FP) \\
\textbf{Compressed File}   & FAIL & true negative (TN) \\
    \bottomrule
  \end{tabular}

  \label{tab:possibleoutcomes}%
\end{table}

After computing the training dataset, we observed that our method could be applied to any file size, due to the stability of the training threshold levels. As supporting evidence, we compute the average of threshold values of our experiments (see Table \ref{tab:thresholdpersize}). One can observe that the average deviation is very low and that the values are very similar for each file size. Therefore, the similarity of the threshold levels obtained from the training tests indicate the stability of the encrypted files' features. The latter means that our threshold values can be applied, regardless of the file-type and file sizes. In addition, we only need to train the method once at the beginning; thus, enabling efficient and adaptable classification.

\section{Discussion}
\label{sec:discussion}

The average accuracy outcomes for each file size according to $\gamma$ are depicted in Figure \ref{fig:accuracy}. In terms of accuracy, the best outcomes are obtained by large file sizes (i.e. 64KB and 32KB). In general, we observe that with a high $\gamma$, we obtain better results because the trade-off between detected compressed files (true positives) and missed encrypted files (false negatives) increases. This is mainly due to the stability of encrypted files' features. Therefore, the use of $\gamma$ enhances the dynamism and adaptability of the system. We can also observe that our method achieves better accuracy than the state-of-the-art (see Figure \ref{fig:accuracy}, 1KB-Others), since the accuracy achieved by the authors in \cite{hahn2018detecting} was 66.9\% and ours is 70.61\%, using the same benchmark dataset proposed by them.

For the sake of clarity, Table \ref{tab:detailedoutcomes} provides the detailed outcomes (as described in Table \ref{tab:possibleoutcomes}) for the most restrictive and the most relaxed case in each file size (i.e. for $\gamma=0.1$ and $\gamma=2$, respectively). We can observe that the values for encrypted files (i.e. TP and FN) are very stable, which means that, as expected, file's randomness properties are similar regardless of the file sizes, according to the tested values (i.e. a study of the randomness properties of files smaller than 1KB is left to future work). In the case of compressed files (i.e. FP and TN) their behaviour is rather variable in the case of small files. Nevertheless, compressed files' outcomes become more stable the higher the file size (i.e. the difference between outcomes for big files is lower than for small files). Moreover, such difference is also reduced when using $\gamma =0.1 $ and $\gamma = 2$ for high file sizes, which means that we can apply more relaxed policies in such cases.
In general, the total amount of correctly classified files (i.e. accuracy) is higher when we apply $\gamma = 2$, since the degrowth rate of the TN value is lower than the growth rate of the TP value.

\begin{table}[tb!]
\small
\renewcommand{\arraystretch}{1.25}
\renewcommand{\tabcolsep}{2mm}
  \centering
   \caption{Excerpt of the outcomes (percentages) for the worst (most restrictive) and best (most relaxed) strategies.}
   \resizebox{\columnwidth}{!}{%
  \begin{tabular}{ccccccc} %
    \toprule
 \multicolumn{1}{c}{\textbf{File size}}   &
 \multicolumn{1}{c}{\textbf{$\gamma$}}  & \multicolumn{1}{c}{\textbf{TP}}  & \multicolumn{1}{c}{\textbf{FN}}& \multicolumn{1}{c}{\textbf{FP}}& \multicolumn{1}{c}{\textbf{TN}} & \multicolumn{1}{c}{\textbf{Accuracy (TP+TN)}}\\
    \midrule

\multirow{2}*{\textbf{1 (Others)}}   & 0.1 & 26.72 &23.28  &13.16  & 36.84 & 63.56\\
  & 2 & 46.70&3.29 &26.09 &23.90 & 70.60 \\ \hline

\multirow{2}*{\textbf{1}}   & 0.1 & 26.73 &23.27  &14.31  & 35.69 & 62.42\\
& 2 & 46.69&3.31& 28.01&  21.99 &68.68 \\ \hline

\multirow{2}*{\textbf{2}}   & 0.1 & 26.58 &23.42  &10.41  & 39.59 & 66.17\\
& 2 &46.68& 3.32&21.18  &  28.82 & 75.50 \\ \hline

\multirow{2}*{\textbf{4}}   & 0.1 & 26.74 &23.26  &7.13  & 42.87 & 69.61 \\
& 2 & 46.77& 3.23&  15.1&34.90 &81.67 \\ \hline

\multirow{2}*{\textbf{8}}   & 0.1 & 26.68 &23.32  &4.84  & 45.16 & 71.84\\
   & 2 & 46.69& 3.31& 10.49& 39.51 & 86.20\\ \hline

\multirow{2}*{\textbf{16}}   & 0.1 & 26.74 &23.26  &3.11  & 46.89 & 73.63 \\
   & 2 &46.86& 3.14& 7.01& 42.99 & 89.85 \\ \hline

\multirow{2}*{\textbf{32}}   & 0.1 & 26.85 &23.15  &1.81  & 48.19 & 75.04 \\
  & 2 & 46.66& 3.34& 3.89&46.11 & 92.77\\ \hline

\multirow{2}*{\textbf{64}}   & 0.1 & 27.01 &22.90  &0.98  & 49.02 & 76.03\\
 & 2 & 46.82& 3.18& 2.09&47.91 & 94.72\\

    \bottomrule
  \end{tabular}

  \label{tab:detailedoutcomes}%
 }
\end{table}



Next, after thoroughly analysing the results, we discovered that the randomness tests applied to compressed files exhibited different behaviours according to the input file-type. More precisely, there is a strong relationship between the randomness of the input file-type and the randomness of the compressed file generated. To document such behaviour, we classify the outcomes of all data streams according to input file-types for all the compressed (see Fig \ref{fig:filetypedetail}) algorithms tested (see Table \ref{tab:compencmethods}). First, Fig. \ref{fig:64kbencryp}, shows the accuracy detection of 64KB encrypted data streams, whose behaviour is similar to other file sizes and therefore we omit the results. It is apparent that when we analyse encrypted data streams we cannot distinguish the input file-type which is a result that complies with the theory. On the contrary, one can observe that compressed binary and compressed text files, as well as compressed video files are more easily classified by our method (achieving 100\% accuracy with binary and text in 64KB file streams) than the rest. This implies that their randomness is much lower than other files, such as PDF or MP3 (i.e. MP3 is already a compressed file, as well as JPG files). This behaviour applies to all file sizes except for files lower than 4KB, where compressed text files become more indistinguishable, but we still find notable differences between each file-type. Such findings are relevant, especially from a security perspective, because discovering the content of compressed data streams enables the proper management when detecting exchange of unexpected/suspicious files (e.g. executables). Again, the result complies with the theory since the output of compression algorithms like Huffman or members of the Lempel-Ziv family highly depends on the statistical properties of the input stream. 

In summary, the results show that we can distinguish compressed from encrypted bit streams accurately in an efficient way, and in the former case we may even determine the content of compressed files with certainty. The accuracy of the proposed methodology is highly dependent on the size of the investigated packets, decreasing as the packet size decreases. Moreover, our threshold-based method achieves higher accuracy (see Figure \ref{fig:comparison}) with more efficiency than the state-of-the-art \cite{hahn2018detecting} ($k$-NN and convolutional Neural networks), since they have a complexity of at least $O(n^2)$ whilst our method has linear cost $O(n)$. Moreover, in most cases we need only to compare one feature (and not $n$) to classify a bit stream (e.g. if the chi-square test fails, the file is discarded and there is no need to compute the rest of tests). In addition, results are reproducible since the variability of threshold values is almost negligible, so that a universal threshold can be considered regardless of data, avoiding costly training procedures (such in the case of Neural networks) and other data-dependent methods.

\section{Conclusions}
\label{sec:conclusions}
Distinguishing high entropy sources (e.g. encrypted) from compressed streams is a challenge that has been the focus of the research community. Most solutions in the literature achieve high accuracy only when they are able to access all the exchanged packets and extract patterns from the intercepted traffic, but are inefficient when trying to distinguish between high entropy data \cite{olivain2006detecting,dorfinger2010entropy,paninski2003estimation,paninski2004estimating,malhotra2007detection,6005446,conti2010automated}. Recently, Hahn et al. \cite{hahn2018detecting} introduced a method based on machine learning that manages to distinguish between encrypted and compressed data using random packets and not their complete sequence.

In this article, we introduced a novel threshold-based methodology that uses efficient randomness tests to classify data streams into either compressed or encrypted by investigating each intercepted packet individually. Our approach was evaluated using a statistically sound benchmark we developed, achieving an accuracy between 68.68\% (worst case scenario) and 94.72\% (and 70.61\% using the reference dataset of \cite{hahn2018detecting}). The accuracy rate depends on the packet size. We also determined that the randomness of compressed files has a strong dependence on the input file-type and we analysed their behaviour. We were able to conclude that (compressed) binary, text and video files can be easily detected by our system. Moreover, our method is more efficient than other competing state-of-the-art works and enables real-time traffic classification.

In the future, we plan to: (i) refine this method to further increase its accuracy; especially for low packet sizes, and (ii) enable more accurate content classification of compressed files, to improve pro-active security and specific file-type detection. Evaluation will also be carried  in real-time environment, such as those of the authors' institutions.

\appendix
\subsection{Randomness Tests}
\label{sec:randomnesstests}
As this work focuses on the randomness analysis of data streams to determine whether they are encrypted, we will now briefly describe widely used randomness evaluation methods in the literature.

\noindent \textbf{Entropy}. According to Information Theory \cite{shannon1949communication}, the entropy measures the unpredictability of data, given an information source. Therefore, the higher the entropy, the higher the randomness of data.

\noindent \textbf{Chi-square test}. This is one of the prevalent tests used to compute the randomness of a data stream. The test computes how much two given data samples differ from each other. It is used to test whether a set of data follows a specific distribution with a degree of confidence. Usually, the chi-square outcomes are represented as an absolute value and a confidence percentage $\chi\%$, which estimates the frequency that a truly random sequence exceeds this value by chance. There are three main possibilities \cite{d2017goodness}:
\begin{enumerate}
    \item  $ 1\%>\chi\%$ or $\chi\%>99\%$, the stream is not random.
    \item $ 1\%< \chi\% <5\% $ or $ 95\%<\chi\%<99\%$, the stream is  ``suspected'' to be random.
    \item $5\%< \chi\% <10\%$ or $90\%< \chi\% <95\%$, the stream is likely not to be random.
\end{enumerate}

\noindent \textbf{Auto-correlation test}. This test computes inner correlations that can reveal a cyclic pattern or periodic behaviours. The randomness of the data is checked by calculating auto-correlation for the values of the data stream at different time lags \cite{ghosh1991handbook}. Auto-correlation values close to zero show a highly random pattern, while high variations from zero show non-random behaviour.

\noindent \textbf{Average test}. This simple test computes the summation of the values of the bytes. For a random data stream, the value should be close to 127.5.

\noindent \textbf{Kolmogorov-Smirnov test}. This test is an alternative to the chi-square test and quantifies the maximum absolute difference between two empirical cumulative distribution functions as a measure of
disagreement \cite{lopes2011kolmogorov}.

\noindent \textbf{Anderson-Darlong test}. This test is similar to Kolmogorov-Smirnov test, however its reliability is higher, especially for small data streams, as reported in \cite{razali2011power}.

\noindent \textbf{Monte Carlo value for $\pi$ test}. This method computes an approximation to the number $\pi$  using the bit values of the data stream \cite{lo1989size}. The closer to $\pi$, the more random is the data stream.

\noindent \textbf{Monobit test}. This test tries to estimate the balance of ones and zeros in a bit stream. Practically, given a sequence of $n$ bits, we test whether:
\[
    erfc(\frac{|\#zeroes-\#ones|}{n\sqrt{2}})<0.01
\]

\noindent \textbf{Poker test}. This test splits the bit stream in segments of 4 bits. Each segment, when converted to an integer belongs to $[0,15]$. If we denote as $f(i)$ the number of occurences of each number $i$ then we evaluate:
\[
    X=\frac{16}{5000}\sum_{i=0}^{15}f(i)^2-5000
\]
The test is passed if $2.16<X<46.17$.

\noindent \textbf{Runs test}. This test evaluates how often we see the same consecutive bits (a run), e.g. all ones in a raw, in a bit stream. To this end, this test we count the length of each possible run up to length 6 and check whether it conforms to a specific pattern.

\noindent \textbf{Long runs test}. In this test we try to determine whether there are runs of length more than 25.

\noindent \textbf{Diehard tests}. The Diehard tests are a collection of statistical and mathematical tests for measuring the quality of a random number generator \cite{marsaglia1993monkey}.

\noindent \textbf{TestU01 tests}. Similarly to Diehard, the TestU01 suite is a collection of random number generator methods as well as a collection of tests to measure their quality \cite{l2007testu01}.

\noindent \textbf{FIPS-2-140 test}. The National Institute of Standards and Technology (NIST) proposed a set of four empirical tests to analyse the randomness of binary data streams \cite{nisttests}. In our experiments, we use the FIPS-2-140 cryptographic module test with minimum block size (i.e. 20,000 bits) and, hence, tests are applied independently to each data block. The set consists of the monobit, the poker, the runs, and the long runs tests presented above.

\noindent \textbf{SP 800-22 test suite}. NIST later extended the FIPS-2-140 test suite for random and pseudorandom number generators with more tests in SP 800-22 \cite{heckert800andrew}.

\section*{Acknowledgments}
This work was supported by the European Commission under the Horizon 2020 Programme (H2020), as part of the project YAKSHA (Grant Agreement no. 780498) and is co‐financed by the European Union and Greek national funds through the Operational Program Competitiveness, Entrepreneurship and Innovation, under the call RESEARCH - CREATE - INNOVATE (project code:T1EDK-01958).

\bibliographystyle{plain}

\end{document}